\documentclass[reprint,superscriptaddress,amsmath,amssymb,aps,longbibliography]{revtex4-1}
\usepackage{graphicx}% Include figure files amsmath,amssymb,aps,floatfix]{revtex4-1}
\pdfoutput=1
\usepackage{epstopdf}
\usepackage{color}
\usepackage{dcolumn}% Align table columns on decimal point
\usepackage{bm}% bold math
\usepackage[colorlinks=true,allcolors=blue]{hyperref}% add hypertext capabilities
\usepackage{upgreek} % for Greek letters that aren't variables
\usepackage{wasysym}
\usepackage{nicefrac}
\usepackage{siunitx}
\usepackage{booktabs}
\usepackage[normalem]{ulem} %STRIKE
\usepackage{threeparttable} % for more complex tables
\usepackage{tabularx}

\begin{document}

\title{Results of a Search for Sub-GeV Dark Matter Using 2013 LUX Data}

\author{D.S.~Akerib} \affiliation{Case Western Reserve University, Department of Physics, 10900 Euclid Avenue, Cleveland, Ohio  44106, USA} \affiliation{SLAC National Accelerator Laboratory, 2575 Sand Hill Road, Menlo Park, California 94205, USA} \affiliation{Kavli Institute for Particle Astrophysics and Cosmology, Stanford University, 452 Lomita Mall, Stanford, California 94309, USA}
\author{S.~Alsum} \affiliation{University of Wisconsin-Madison, Department of Physics, 1150 University Avenue, Madison, Wisconsin  53706, USA}  
\author{H.M.~Ara\'{u}jo} \affiliation{Imperial College London, High Energy Physics, Blackett Laboratory, London SW7 2BZ, United Kingdom}  
\author{X.~Bai} \affiliation{South Dakota School of Mines and Technology, 501 East St Joseph Street, Rapid City, South Dakota   57701, USA}  
%\author{A.J.~Bailey} \affiliation{Imperial College London, High Energy Physics, Blackett Laboratory, London SW7 2BZ, United Kingdom}  
\author{J.~Balajthy} \affiliation{University of California Davis, Department of Physics, One Shields Avenue, Davis, California 95616, USA}  
%\author{A.~Baxter} \affiliation{University of Liverpool, Department of Physics, Liverpool L69 7ZE, United Kingdom}  
\author{P.~Beltrame} \affiliation{SUPA, School of Physics and Astronomy, University of Edinburgh, Edinburgh EH9 3FD, United Kingdom}  
\author{E.P.~Bernard} \affiliation{University of California Berkeley, Department of Physics, Berkeley, California 94720, USA}  
\author{A.~Bernstein} \affiliation{Lawrence Livermore National Laboratory, 7000 East Avenue, Livermore, California 94551, USA}  
\author{T.P.~Biesiadzinski} \affiliation{Case Western Reserve University, Department of Physics, 10900 Euclid Avenue, Cleveland, Ohio  44106, USA} \affiliation{SLAC National Accelerator Laboratory, 2575 Sand Hill Road, Menlo Park, California 94205, USA} \affiliation{Kavli Institute for Particle Astrophysics and Cosmology, Stanford University, 452 Lomita Mall, Stanford, California 94309, USA}
\author{E.M.~Boulton} \affiliation{University of California Berkeley, Department of Physics, Berkeley, California 94720, USA} \affiliation{Lawrence Berkeley National Laboratory, 1 Cyclotron Road, Berkeley, California 94720, USA} \affiliation{Yale University, Department of Physics, 217 Prospect Street, New Haven, CT 06511, USA}
\author{B.~Boxer} \affiliation{University of Liverpool, Department of Physics, Liverpool L69 7ZE, United Kingdom}  
\author{P.~Br\'as} \affiliation{LIP-Coimbra, Department of Physics, University of Coimbra, Rua Larga, 3004-516 Coimbra, Portugal}  
\author{S.~Burdin} \affiliation{University of Liverpool, Department of Physics, Liverpool L69 7ZE, United Kingdom}  
\author{D.~Byram} \affiliation{University of South Dakota, Department of Physics, 414E Clark Street, Vermillion, South Dakota   57069, USA} \affiliation{South Dakota Science and Technology Authority, Sanford Underground Research Facility, Lead, South Dakota   57754, USA} 
%\author{S.B.~Cahn} \affiliation{Yale University, Department of Physics, 217 Prospect Street, New Haven, CT 06511, USA}  
\author{M.C.~Carmona-Benitez} \affiliation{Pennsylvania State University, Department of Physics, 104 Davey Lab, University Park, Pennsylvania    16802-6300, USA}  
\author{C.~Chan} \affiliation{Brown University, Department of Physics, 182 Hope Street, Providence, Rhode Island   02912, USA}  
%\author{A.A.~Chiller} \affiliation{University of South Dakota, Department of Physics, 414E Clark Street, Vermillion, South Dakota   57069, USA}  
%\author{C.~Chiller} \affiliation{University of South Dakota, Department of Physics, 414E Clark Street, Vermillion, South Dakota   57069, USA}  
%\author{A.~Currie} \affiliation{Imperial College London, High Energy Physics, Blackett Laboratory, London SW7 2BZ, United Kingdom}  
\author{J.E.~Cutter} \affiliation{University of California Davis, Department of Physics, One Shields Avenue, Davis, California 95616, USA}  
\author{T.J.R.~Davison} \affiliation{SUPA, School of Physics and Astronomy, University of Edinburgh, Edinburgh EH9 3FD, United Kingdom}  
%\author{L.~de\,Viveiros}  \affiliation{Pennsylvania State University, Department of Physics, 104 Davey Lab, University Park, Pennsylvania    16802-6300, USA}  
%\author{A.~Dobi} \affiliation{Lawrence Berkeley National Laboratory, 1 Cyclotron Road, Berkeley, California 94720, USA}  
%\author{J.E.Y.~Dobson} \affiliation{Department of Physics and Astronomy, University College London, Gower Street, London WC1E 6BT, United Kingdom}  
\author{E.~Druszkiewicz} \affiliation{University of Rochester, Department of Physics and Astronomy, Rochester, New York 14627, USA}  
%\author{B.N.~Edwards} \affiliation{Yale University, Department of Physics, 217 Prospect Street, New Haven, CT 06511, USA}  
%\author{C.H.~Faham} \affiliation{Lawrence Berkeley National Laboratory, 1 Cyclotron Road, Berkeley, California 94720, USA}  
\author{S.R.~Fallon} \affiliation{University at Albany, State University of New York, Department of Physics, 1400 Washington Avenue, Albany, New York 12222, USA}  
\author{A.~Fan} \affiliation{SLAC National Accelerator Laboratory, 2575 Sand Hill Road, Menlo Park, California 94205, USA} \affiliation{Kavli Institute for Particle Astrophysics and Cosmology, Stanford University, 452 Lomita Mall, Stanford, California 94309, USA} 
\author{S.~Fiorucci} \affiliation{Lawrence Berkeley National Laboratory, 1 Cyclotron Road, Berkeley, California 94720, USA} \affiliation{Brown University, Department of Physics, 182 Hope Street, Providence, Rhode Island   02912, USA} 
\author{R.J.~Gaitskell} \affiliation{Brown University, Department of Physics, 182 Hope Street, Providence, Rhode Island   02912, USA}  
%\author{V.M.~Gehman} \affiliation{Lawrence Berkeley National Laboratory, 1 Cyclotron Road, Berkeley, California 94720, USA}  
\author{J.~Genovesi} \affiliation{University at Albany, State University of New York, Department of Physics, 1400 Washington Avenue, Albany, New York 12222, USA}  
\author{C.~Ghag} \affiliation{Department of Physics and Astronomy, University College London, Gower Street, London WC1E 6BT, United Kingdom}  
%\author{K.R.~Gibson} \affiliation{Case Western Reserve University, Department of Physics, 10900 Euclid Avenue, Cleveland, Ohio  44106, USA}  
\author{M.G.D.~Gilchriese} \affiliation{Lawrence Berkeley National Laboratory, 1 Cyclotron Road, Berkeley, California 94720, USA}  
%\author{E.~Grace} \affiliation{Pennsylvania State University, Department of Physics, 104 Davey Lab, University Park, Pennsylvania    16802-6300, USA}  
\author{C.~Gwilliam} \affiliation{University of Liverpool, Department of Physics, Liverpool L69 7ZE, United Kingdom}  
\author{C.R.~Hall} \affiliation{University of Maryland, Department of Physics, College Park, MD 20742, USA}  
%\author{M.~Hanhardt} \affiliation{South Dakota School of Mines and Technology, 501 East St Joseph Street, Rapid City, South Dakota   57701, USA} \affiliation{South Dakota Science and Technology Authority, Sanford Underground Research Facility, Lead, South Dakota   57754, USA} 
\author{S.J.~Haselschwardt} \affiliation{University of California Santa Barbara, Department of Physics, Santa Barbara, California 93106, USA}  
\author{S.A.~Hertel} \affiliation{University of Massachusetts, Amherst Center for Fundamental Interactions and Department of Physics, Amherst, Massachusetts   01003-9337 USA} \affiliation{Lawrence Berkeley National Laboratory, 1 Cyclotron Road, Berkeley, California 94720, USA} 
\author{D.P.~Hogan} \affiliation{University of California Berkeley, Department of Physics, Berkeley, California 94720, USA}  
\author{M.~Horn} \affiliation{South Dakota Science and Technology Authority, Sanford Underground Research Facility, Lead, South Dakota   57754, USA} \affiliation{University of California Berkeley, Department of Physics, Berkeley, California 94720, USA} 
\author{D.Q.~Huang} \affiliation{Brown University, Department of Physics, 182 Hope Street, Providence, Rhode Island   02912, USA}  
\author{C.M.~Ignarra} \affiliation{SLAC National Accelerator Laboratory, 2575 Sand Hill Road, Menlo Park, California 94205, USA} \affiliation{Kavli Institute for Particle Astrophysics and Cosmology, Stanford University, 452 Lomita Mall, Stanford, California 94309, USA} 
\author{R.G.~Jacobsen} \affiliation{University of California Berkeley, Department of Physics, Berkeley, California 94720, USA}  
\author{O.~Jahangir} \affiliation{Department of Physics and Astronomy, University College London, Gower Street, London WC1E 6BT, United Kingdom}  
\author{W.~Ji} \affiliation{Case Western Reserve University, Department of Physics, 10900 Euclid Avenue, Cleveland, Ohio  44106, USA} \affiliation{SLAC National Accelerator Laboratory, 2575 Sand Hill Road, Menlo Park, California 94205, USA} \affiliation{Kavli Institute for Particle Astrophysics and Cosmology, Stanford University, 452 Lomita Mall, Stanford, California 94309, USA}
\author{K.~Kamdin} \affiliation{University of California Berkeley, Department of Physics, Berkeley, California 94720, USA} \affiliation{Lawrence Berkeley National Laboratory, 1 Cyclotron Road, Berkeley, California 94720, USA} 
\author{K.~Kazkaz} \affiliation{Lawrence Livermore National Laboratory, 7000 East Avenue, Livermore, California 94551, USA}  
\author{D.~Khaitan} \affiliation{University of Rochester, Department of Physics and Astronomy, Rochester, New York 14627, USA}  
\author{R.~Knoche} \affiliation{University of Maryland, Department of Physics, College Park, MD 20742, USA}  
\author{E.V.~Korolkova} \affiliation{University of Sheffield, Department of Physics and Astronomy, Sheffield, S3 7RH, United Kingdom}  
\author{S.~Kravitz} \affiliation{Lawrence Berkeley National Laboratory, 1 Cyclotron Road, Berkeley, California 94720, USA}  
\author{V.A.~Kudryavtsev} \affiliation{University of Sheffield, Department of Physics and Astronomy, Sheffield, S3 7RH, United Kingdom}  
%\author{N.A.~Larsen} \affiliation{Yale University, Department of Physics, 217 Prospect Street, New Haven, CT 06511, USA}  
%\author{C.~Lee} \affiliation{Case Western Reserve University, Department of Physics, 10900 Euclid Avenue, Cleveland, Ohio  44106, USA} \affiliation{SLAC National Accelerator Laboratory, 2575 Sand Hill Road, Menlo Park, California 94205, USA} \affiliation{Kavli Institute for Particle Astrophysics and Cosmology, Stanford University, 452 Lomita Mall, Stanford, California 94309, USA}
\author{B.G.~Lenardo} \affiliation{University of California Davis, Department of Physics, One Shields Avenue, Davis, California 95616, USA} \affiliation{Lawrence Livermore National Laboratory, 7000 East Avenue, Livermore, California 94551, USA} 
\author{K.T.~Lesko} \affiliation{Lawrence Berkeley National Laboratory, 1 Cyclotron Road, Berkeley, California 94720, USA}  
%\author{C.~Levy} \affiliation{University at Albany, State University of New York, Department of Physics, 1400 Washington Avenue, Albany, New York 12222, USA} \affiliation{Lawrence Berkeley National Laboratory, 1 Cyclotron Road, Berkeley, California 94720, USA} 
\author{J.~Liao} \affiliation{Brown University, Department of Physics, 182 Hope Street, Providence, Rhode Island   02912, USA}  
\author{J.~Lin} \affiliation{University of California Berkeley, Department of Physics, Berkeley, California 94720, USA}  
\author{A.~Lindote} \affiliation{LIP-Coimbra, Department of Physics, University of Coimbra, Rua Larga, 3004-516 Coimbra, Portugal}  
\author{M.I.~Lopes} \affiliation{LIP-Coimbra, Department of Physics, University of Coimbra, Rua Larga, 3004-516 Coimbra, Portugal}  
\author{A.~Manalaysay} \affiliation{University of California Davis, Department of Physics, One Shields Avenue, Davis, California 95616, USA}  
\author{R.L.~Mannino} \affiliation{Texas A \& M University, Department of Physics, College Station, Texas 77843, USA} \affiliation{University of Wisconsin-Madison, Department of Physics, 1150 University Avenue, Madison, Wisconsin  53706, USA} 
\author{N.~Marangou} \affiliation{Imperial College London, High Energy Physics, Blackett Laboratory, London SW7 2BZ, United Kingdom}  
\author{M.F.~Marzioni} \affiliation{SUPA, School of Physics and Astronomy, University of Edinburgh, Edinburgh EH9 3FD, United Kingdom}  
\author{D.N.~McKinsey} \affiliation{University of California Berkeley, Department of Physics, Berkeley, California 94720, USA} \affiliation{Lawrence Berkeley National Laboratory, 1 Cyclotron Road, Berkeley, California 94720, USA} 
\author{D.-M.~Mei} \affiliation{University of South Dakota, Department of Physics, 414E Clark Street, Vermillion, South Dakota   57069, USA}  
%\author{J.~Mock} \affiliation{University at Albany, State University of New York, Department of Physics, 1400 Washington Avenue, Albany, New York 12222, USA}  
\author{M.~Moongweluwan} \affiliation{University of Rochester, Department of Physics and Astronomy, Rochester, New York 14627, USA}  
\author{J.A.~Morad} \affiliation{University of California Davis, Department of Physics, One Shields Avenue, Davis, California 95616, USA}  
\author{A.StreetJ.~Murphy} \affiliation{SUPA, School of Physics and Astronomy, University of Edinburgh, Edinburgh EH9 3FD, United Kingdom}  
\author{A.~Naylor} \affiliation{University of Sheffield, Department of Physics and Astronomy, Sheffield, S3 7RH, United Kingdom}  
\author{C.~Nehrkorn} \affiliation{University of California Santa Barbara, Department of Physics, Santa Barbara, California 93106, USA}  
\author{H.N.~Nelson} \affiliation{University of California Santa Barbara, Department of Physics, Santa Barbara, California 93106, USA}  
\author{F.~Neves} \affiliation{LIP-Coimbra, Department of Physics, University of Coimbra, Rua Larga, 3004-516 Coimbra, Portugal}  
%\author{K.~O'Sullivan} \affiliation{University of California Berkeley, Department of Physics, Berkeley, California 94720, USA} \affiliation{Lawrence Berkeley National Laboratory, 1 Cyclotron Road, Berkeley, California 94720, USA} \affiliation{Yale University, Department of Physics, 217 Prospect Street, New Haven, CT 06511, USA}
\author{K.C.~Oliver-Mallory} \affiliation{University of California Berkeley, Department of Physics, Berkeley, California 94720, USA} \affiliation{Lawrence Berkeley National Laboratory, 1 Cyclotron Road, Berkeley, California 94720, USA} 
\author{K.J.~Palladino} \affiliation{University of Wisconsin-Madison, Department of Physics, 1150 University Avenue, Madison, Wisconsin  53706, USA}  
\author{E.K.~Pease} \affiliation{University of California Berkeley, Department of Physics, Berkeley, California 94720, USA} \affiliation{Lawrence Berkeley National Laboratory, 1 Cyclotron Road, Berkeley, California 94720, USA} 
%\author{L.~Reichhart} \affiliation{Department of Physics and Astronomy, University College London, Gower Street, London WC1E 6BT, United Kingdom}  
\author{Q.~Riffard} \affiliation{University of California Berkeley, Department of Physics, Berkeley, California 94720, USA} \affiliation{Lawrence Berkeley National Laboratory, 1 Cyclotron Road, Berkeley, California 94720, USA} 
\author{G.R.C.~Rischbieter} \affiliation{University at Albany, State University of New York, Department of Physics, 1400 Washington Avenue, Albany, New York 12222, USA}  
\author{C.~Rhyne} \affiliation{Brown University, Department of Physics, 182 Hope Street, Providence, Rhode Island   02912, USA}  
\author{P.~Rossiter} \affiliation{University of Sheffield, Department of Physics and Astronomy, Sheffield, S3 7RH, United Kingdom}  
\author{S.~Shaw} \affiliation{University of California Santa Barbara, Department of Physics, Santa Barbara, California 93106, USA} \affiliation{Department of Physics and Astronomy, University College London, Gower Street, London WC1E 6BT, United Kingdom} 
\author{T.A.~Shutt} \affiliation{Case Western Reserve University, Department of Physics, 10900 Euclid Avenue, Cleveland, Ohio  44106, USA} \affiliation{SLAC National Accelerator Laboratory, 2575 Sand Hill Road, Menlo Park, California 94205, USA} \affiliation{Kavli Institute for Particle Astrophysics and Cosmology, Stanford University, 452 Lomita Mall, Stanford, California 94309, USA}
\author{C.~Silva} \affiliation{LIP-Coimbra, Department of Physics, University of Coimbra, Rua Larga, 3004-516 Coimbra, Portugal}  
\author{M.~Solmaz} \affiliation{University of California Santa Barbara, Department of Physics, Santa Barbara, California 93106, USA}  
\author{V.N.~Solovov} \affiliation{LIP-Coimbra, Department of Physics, University of Coimbra, Rua Larga, 3004-516 Coimbra, Portugal}  
\author{P.~Sorensen} \affiliation{Lawrence Berkeley National Laboratory, 1 Cyclotron Road, Berkeley, California 94720, USA}  
%\author{S.~Stephenson} \affiliation{University of California Davis, Department of Physics, One Shields Avenue, Davis, California 95616, USA}  
\author{T.J.~Sumner} \affiliation{Imperial College London, High Energy Physics, Blackett Laboratory, London SW7 2BZ, United Kingdom}  
\author{M.~Szydagis} \affiliation{University at Albany, State University of New York, Department of Physics, 1400 Washington Avenue, Albany, New York 12222, USA}  
\author{D.J.~Taylor} \affiliation{South Dakota Science and Technology Authority, Sanford Underground Research Facility, Lead, South Dakota   57754, USA}  
\author{W.C.~Taylor} \affiliation{Brown University, Department of Physics, 182 Hope Street, Providence, Rhode Island   02912, USA}  
\author{B.P.~Tennyson} \affiliation{Yale University, Department of Physics, 217 Prospect Street, New Haven, CT 06511, USA}  
\author{P.A.~Terman} \affiliation{Texas A \& M University, Department of Physics, College Station, Texas 77843, USA}  
\author{D.R.~Tiedt} \affiliation{South Dakota School of Mines and Technology, 501 East St Joseph Street, Rapid City, South Dakota   57701, USA}  
\author{W.H.~To} \affiliation{California State University Stanislaus, Department of Physics, 1 University Circle, Turlock, California 95382, USA}  
\author{M.~Tripathi} \affiliation{University of California Davis, Department of Physics, One Shields Avenue, Davis, California 95616, USA}  
\author{L.~Tvrznikova} \email{lucie.tvrznikova@yale.edu}
\affiliation{University of California Berkeley, Department of Physics, Berkeley, California 94720, USA} \affiliation{Lawrence Berkeley National Laboratory, 1 Cyclotron Road, Berkeley, California 94720, USA} \affiliation{Yale University, Department of Physics, 217 Prospect Street, New Haven, CT 06511, USA}
\author{U.~Utku} \affiliation{Department of Physics and Astronomy, University College London, Gower Street, London WC1E 6BT, United Kingdom}  
\author{S.~Uvarov} \affiliation{University of California Davis, Department of Physics, One Shields Avenue, Davis, California 95616, USA}  
\author{V.~Velan} \affiliation{University of California Berkeley, Department of Physics, Berkeley, California 94720, USA}  
%\author{J.R.~Verbus} \affiliation{Brown University, Department of Physics, 182 Hope Street, Providence, Rhode Island   02912, USA}  
\author{R.C.~Webb} \affiliation{Texas A \& M University, Department of Physics, College Station, Texas 77843, USA}  
\author{J.T.~White} \affiliation{Texas A \& M University, Department of Physics, College Station, Texas 77843, USA}  
\author{T.J.~Whitis} \affiliation{Case Western Reserve University, Department of Physics, 10900 Euclid Avenue, Cleveland, Ohio  44106, USA} \affiliation{SLAC National Accelerator Laboratory, 2575 Sand Hill Road, Menlo Park, California 94205, USA} \affiliation{Kavli Institute for Particle Astrophysics and Cosmology, Stanford University, 452 Lomita Mall, Stanford, California 94309, USA}
\author{M.S.~Witherell} \affiliation{Lawrence Berkeley National Laboratory, 1 Cyclotron Road, Berkeley, California 94720, USA}  
\author{F.L.H.~Wolfs} \affiliation{University of Rochester, Department of Physics and Astronomy, Rochester, New York 14627, USA}  
\author{D.~Woodward} \affiliation{Pennsylvania State University, Department of Physics, 104 Davey Lab, University Park, Pennsylvania    16802-6300, USA}  
\author{J.~Xu} \affiliation{Lawrence Livermore National Laboratory, 7000 East Avenue, Livermore, California 94551, USA}  
\author{K.~Yazdani} \affiliation{Imperial College London, High Energy Physics, Blackett Laboratory, London SW7 2BZ, United Kingdom}  
%\author{S.K.~Young} \affiliation{University at Albany, State University of New York, Department of Physics, 1400 Washington Avenue, Albany, New York 12222, USA}  
\author{C.~Zhang} \affiliation{University of South Dakota, Department of Physics, 414E Clark Street, Vermillion, South Dakota   57069, USA}

\collaboration{LUX Collaboration}

\date{\today}% any date may be explicitly specified
\vspace{10 mm}
\begin{abstract}
%\vspace{13 mm}
The scattering of dark matter (DM) particles with sub-GeV masses off nuclei is difficult to detect using liquid xenon-based DM search instruments because the energy transfer during nuclear recoils is smaller than the typical detector threshold. However, the tree-level DM-nucleus scattering diagram can be accompanied by simultaneous emission of a Bremsstrahlung photon or a so-called ``Migdal'' electron. These provide an electron recoil component to the experimental signature at higher energies than the corresponding nuclear recoil. The presence of this signature allows liquid xenon detectors to use both the scintillation and the ionization signals in the analysis where the nuclear recoil signal would not be otherwise visible. We report constraints on spin-independent DM-nucleon scattering for DM particles with masses of 0.4-5~GeV/c$^2$ using 1.4$\times10^4$~kg$\cdot$day of search exposure from the 2013 data from the Large Underground Xenon (LUX) experiment for four different classes of mediators. This analysis extends the reach of liquid xenon-based DM search instruments to lower DM masses than has been achieved previously.
\end{abstract}

\maketitle

\textit{Introduction.---}The two-phase xenon time projection chamber (TPC) is the leading technology used to search for the weakly interacting massive particle (WIMP), a favored dark matter (DM) candidate, in the 5~GeV/c$^2$ to 10~TeV/c$^2$ mass range. Despite substantial improvements in sensitivity over recent years, detecting DM remains an elusive goal~\cite{Akerib:2016vxi,Cui:2017nnn,Aprile:2017iyp}. Consistent progress in ruling out WIMP parameter space has resulted in a significant broadening of efforts, including focusing on lighter particles scattering off nuclei as possible DM candidates. Currently, the intrinsic scintillation properties of nuclear recoils prevent liquid xenon TPCs from reaching sub-GeV DM masses.

Recently, Refs.~\cite{Kouvaris:2016afs, Ibe:2017yqa} proposed novel direct detection channels that extend the reach of liquid xenon detectors to sub-GeV masses. They suggest that DM-nucleus scattering can be accompanied by a signal that results in an electron recoil (ER) at higher energy than the corresponding nuclear recoil (NR) in liquid xenon detectors. Since at low energies ERs produce a stronger signal than NRs, this newly recognized channel enables liquid xenon detectors to reach sub-GeV DM masses. In the Large Underground Xenon (LUX) detector the 50\% detection efficiency for NRs is at 3.3~keV~\cite{Akerib:2015rjg}, compared with 1.24~keV for ERs~\cite{Akerib:2015wdi}. 

This Letter discusses searches of sub-GeV DM in the LUX detector using two different mechanisms: Bremsstrahlung, first proposed in~\cite{Kouvaris:2016afs}, and the Migdal effect, reformulated in~\cite{Ibe:2017yqa}. These atomic inelastic signals are much stronger compared to the traditional elastic NR signal for DM candidates with masses below $\sim5$~GeV/c$^2$. 

Bremsstrahlung considers the emission of a photon from the recoiling atomic nucleus. In the atomic picture, the process can be viewed as the dipole emission of a photon from a xenon atom polarized in the DM-nucleus scattering. The theoretical motivation and event rates for Bremsstrahlung have been derived in~\cite{Kouvaris:2016afs}.

For NRs in liquid xenon, it is usually assumed that electrons around the recoiling nucleus immediately follow the motion of the nucleus so that the atom remains neutral. In reality, the electrons may lag resulting in ionization and excitation of the atom~\cite{Ibe:2017yqa}. When Migdal originally formulated the Migdal effect in 1941~\cite{migdal1941migdal}, he assumed an impulsive force to describe this effect. However, Ref.~\cite{Ibe:2017yqa} reformulated the approach using atomic energy eigenstates for their calculation, thus avoiding the need to resolve the complex time evolution of the nucleus-electron system. Reference~\cite{Ibe:2017yqa} contains the theoretical motivation and presents the expected event rates for the Migdal effect. This analysis conservatively does not consider contributions from the xenon valence electrons ($n=5$), since the surrounding atoms in the liquid may influence the ionization spectrum from these electrons. Contributions from the $n=1,2$ electron shells are negligible at DM masses considered in this study and were also omitted. Furthermore, only electron energy injections caused by ionization were included in the signal model since excitation probabilities are much smaller. 

It should be emphasized that both NR and ER signals are present when considering the Bremsstrahlung and Migdal effects. However, only the ER signal is used in this analysis. The distance traveled by the photon or electron will be less than the position resolution of the detector, always resulting in a single S2. Higher interaction rates in the region of interest are expected from the Migdal effect. 

Both scalar and vector mediators are investigated. The scalar mediator couples to Standard Model (SM) particles by mixing with the SM Higgs boson, and therefore its coupling is proportional to $A^2$, where $A$ is the atomic mass number. The vector mediator considered here, the so-called dark photon, couples to SM particles via mixing with the SM photon, so its coupling is proportional to $Z^2$ where $Z$ is the charge number~\cite{Dolan:2017xbu}. 

Additionally, both heavy and light mediators were studied, motivated by the many hidden (dark) sector DM models~\cite{Ren:2018gyx,Battaglieri:2017aum}. The DM form factor $F_\mathrm{med}(E_R)$ depends on the mass of the particle mediating the interaction at a given recoil energy. For a heavy mediator with $m_\mathrm{med}\gg q$, where $q$ is the momentum transfer, $F_\mathrm{med}$ can be approximated as~1. A heavy scalar mediator is typically assumed for the spin-independent (SI) elastic DM-nucleon cross section~\cite{Lewin:1995rx}. In the light mediator limit, $m_\mathrm{med}\ll q$ and $F_{med}=q_\mathrm{ref} ^4/q^4$, where the SI DM-nucleon cross section is defined at a reference value $q$. For this analysis $q=1$~MeV, a value typical for $m_\mathrm{DM}\lesssim 1$~GeV/c$^2$~\cite{McCabe:2017rln}. Overall, this results in up to four different limits each for the Bremsstrahlung and Migdal signals.

\textit{Data analysis in LUX.---}LUX is a dual-phase (liquid-gas) xenon TPC containing 250~kg of ultrapure liquid xenon in the active detector volume. Energy deposited by a particle interaction in the liquid induces two measurable signals: the prompt primary scintillation signal from VUV photons ($\mathcal{S}$1), and ionization charge. An applied electric field of 180~V/cm drifts these liberated electrons to the surface of the liquid, where the electrons are extracted into the gas and accelerated by a larger electric field, producing secondary electroluminescence photons ($\mathcal{S}$2). Photons are detected by top and bottom arrays with 61 photomultiplier tubes (PMTs) each. The PMT signals from both light pulses, $\mathcal{S}$1 and $\mathcal{S}$2, enable the reconstruction of interaction vertices in three dimensions~\cite{Akerib:2017riv}. The ability to reconstruct positions of interactions in three dimensions allows fiducialization of the active volume. This avoids higher background regions near the detector walls and enables rejection of neutrons and $\gamma$-rays that scatter multiple times within the active detector volume. Furthermore, the ratio of the $\mathcal{S}$1 and $\mathcal{S}$2 signals is exploited to discriminate between ERs and NRs. Details regarding the construction and performance of the LUX detector can be found in~\cite{Akerib:2012ak}.

LUX collected data during two exposures in 2013~\cite{Akerib:2013tjd,Akerib:2015rjg} and from 2014-16~\cite{Akerib:2016vxi}. The work presented here employs WIMP search data with a total exposure of 95 live days using 118~kg of liquid xenon in the fiducial volume collected from April 24 to September 1, 2013, referred to as WS2013. These data have also been used to set limits on spin-dependent interactions~\cite{Akerib:2016lao} and for axion and axionlike particle searches~\cite{Akerib:2017uem}. The performance of the detector during WS2013 is documented in~\cite{Akerib:2017vbi}; only especially relevant information is included here.

Data presented here are identical to the final data set presented in~\cite{Akerib:2015rjg}. Only single scatter events (one $\mathcal{S}$1 followed by one $\mathcal{S}$2) are considered. The fiducial volume is defined from 38-305~$\mu$s in drift time (48.6-8.5~cm above the faces of the bottom PMTs in $z$) and a radius $<20$~cm. $\mathcal{S}$1 pulses are required to have a two-PMT coincidence and produce 1-50~detected photons (phd)~\cite{Faham:2015kqa}. The italicized quantities $S1$ and $S2$ indicate signal amplitudes that have been corrected for geometrical effects and time-dependent xenon purity. Therefore, $S1$ can be below 2.0~phd even when the twofold photon coincidence is satisfied, as discussed in~\cite{Akerib:2017vbi}. A threshold of 165~phd raw $S2$ size is applied to mitigate random coincidence background from small, isolated $S2$s. 

The total energy deposition $E$ of ERs in the detector is directly proportional to the number of quanta produced:
\begin{align}\label{my_first_eqn}
	E=W\left(n_{\gamma} + n_e\right)=W\left(\frac{S1}{g_1}+\frac{S2}{g_2}\right), \nonumber
\end{align}
where $n_{\gamma}$ is the number of photons and $n_e$ the initial number of electrons leaving the interaction site. The detector-specific gain factors $g_1=0.117$~phd per photon and $g_2=12.2$~phd per electron were obtained from calibrations~\cite{Akerib:2017vbi}. The efficiency for extracting electrons from liquid to gas is $49\%\pm3\%$. The overall photon detection efficiency for prompt scintillation, $g_1$, is the product of the average light collection efficiency of the detector and the average PMT quantum efficiency. The corresponding quantity for $S2$ light, $g_2$, consists of the product of the electron extraction efficiency (from liquid to gas) and the average single electron pulse size. The average energy needed to produce a single photon or electron $W$ has a value of $(13.7\pm0.2)$~eV/quanta~\cite{dahl2009physics}.

\begin{figure}[b!]
\includegraphics[width=85mm]{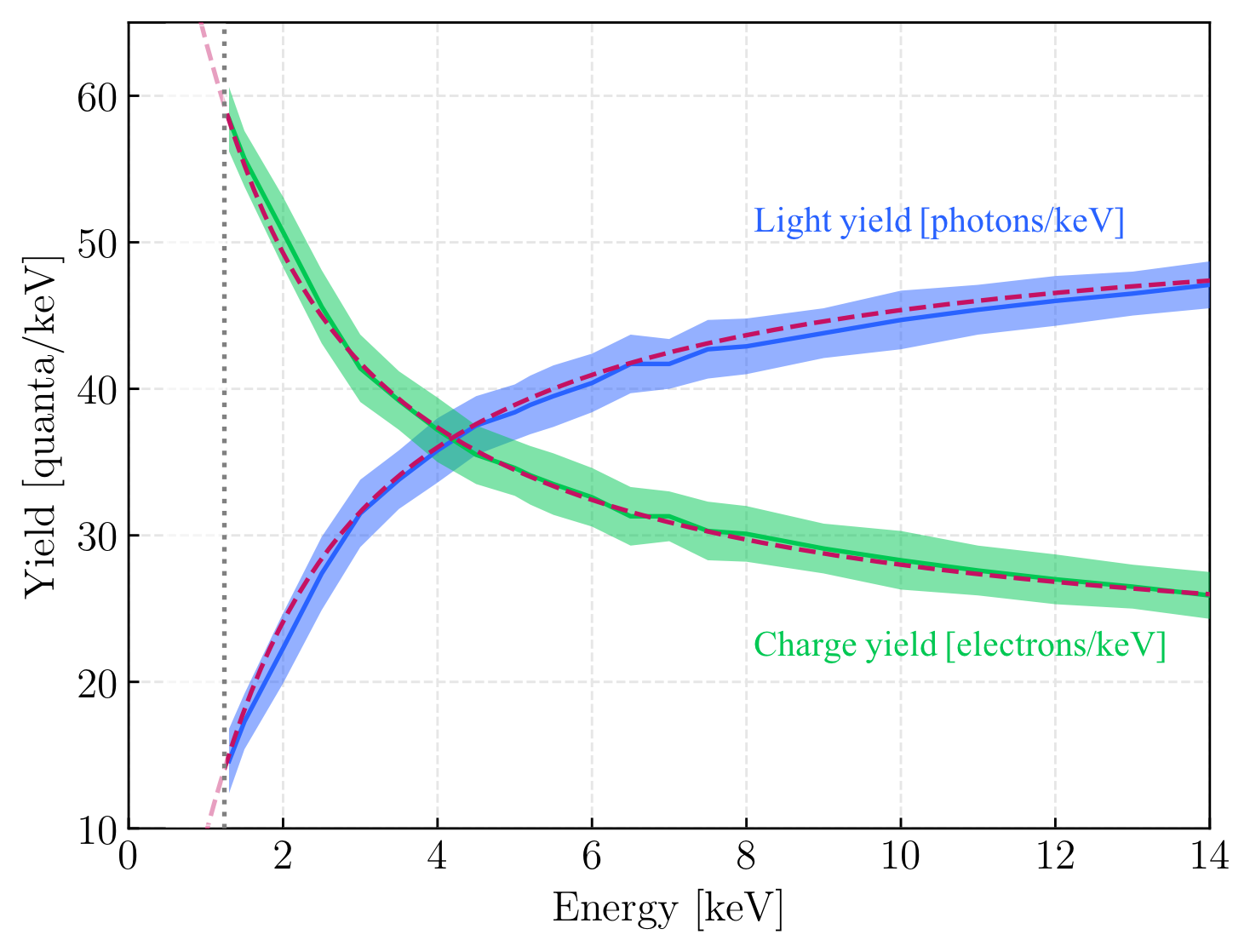}
\caption{The light (blue) and charge (green) yields of tritium ER events as a function of recoil energy as measured \textit{in situ} by the LUX detector at 180~V/cm (solid lines) compared to NEST~v2.0 simulations (dashed pink line). The bands indicate the 1-$\sigma$ systematic uncertainties of the measurement. The dotted gray line shows the 1.24~keV energy threshold implemented in the analysis.}
\label{fig:light_charge_yields}
\end{figure}

\textit{Electron recoil signal yields.---}The response of the LUX detector to ERs was characterized using  internal tritium calibrations performed in December 2013, directly following WS2013. Tritiated methane was injected into the gas circulation to achieve a spatially uniform distribution of events dissolved in the detector's active region, as described in~\cite{Akerib:2015wdi}. This direct calibration is applied to build the signal model for this analysis. Figure~\ref{fig:light_charge_yields} shows excellent agreement between the ER yields from the \textit{in situ} tritium calibrations and yields obtained from the Noble Element Simulation Technique (NEST) package v2.0~\cite{nest2.0}, used to model the ER response in the signal model. The complementary behavior between the light and charge yields is due to recombination effects described in~\cite{Akerib:2016qlr, Akerib:2015wdi}. Since this Letter considers recoils at the lowest energies, where recombination is small, it is limited by light production rather than charge yields.

\begin{figure*}[t!]
\includegraphics[width=130mm]{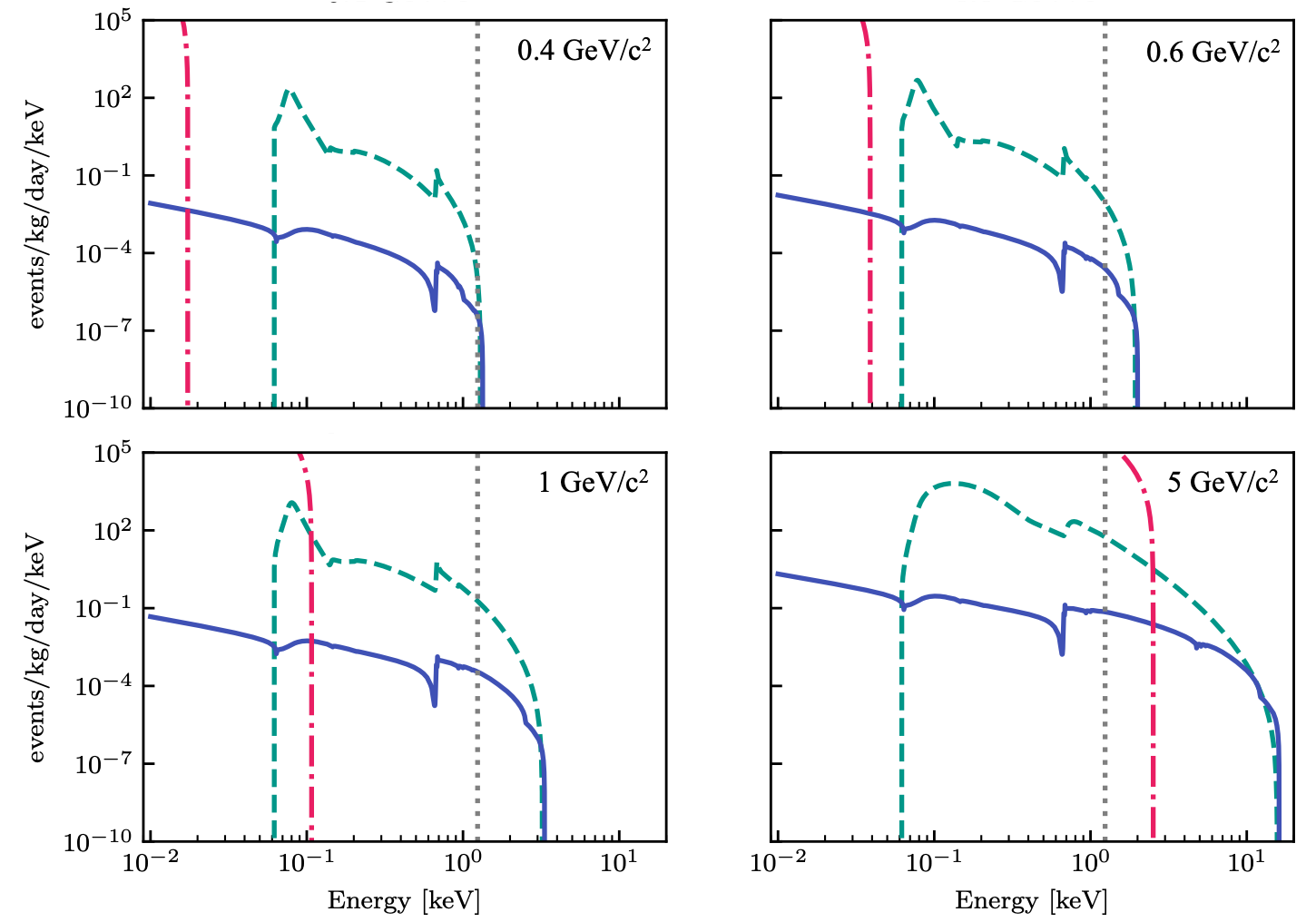}
\caption{Scattering rates in xenon for the Bremsstrahlung (solid blue) and Migdal effects (dashed teal). The DM-nucleus scattering rates resulting in elastic NR in LUX are also shown (dash-dot pink). Also shown is a signal cut off at~1.24 keV (dotted gray) applied in the analysis, corresponding to 50\% efficiency of ER detection. Note that 50\% efficiency for NR event detection occurs at 3.3~keV~\cite{Akerib:2015rjg}.}
\label{fig:Brem_Migdal}
\end{figure*}

A 1.24~keV low-energy cutoff was applied in the signal model corresponding to 50\% efficiency of ER detection (cf. Fig.~6 in~\cite{Akerib:2015wdi}), which imposes a lower mass limit on DM sensitivity of 0.4~GeV/c$^2$. The highest tested mass was chosen to be 5~GeV/c$^2$ since at higher masses the traditional elastic NR results in a larger event rate above threshold than the Bremsstrahlung or Migdal effects. The scattering rates for both the Bremsstrahlung and Migdal effects along with the traditional elastic NR signal and the impact of the signal cutoff for several DM masses are illustrated in Fig.~\ref{fig:Brem_Migdal}.

\begin{figure}[b!]
\includegraphics[width=85mm]{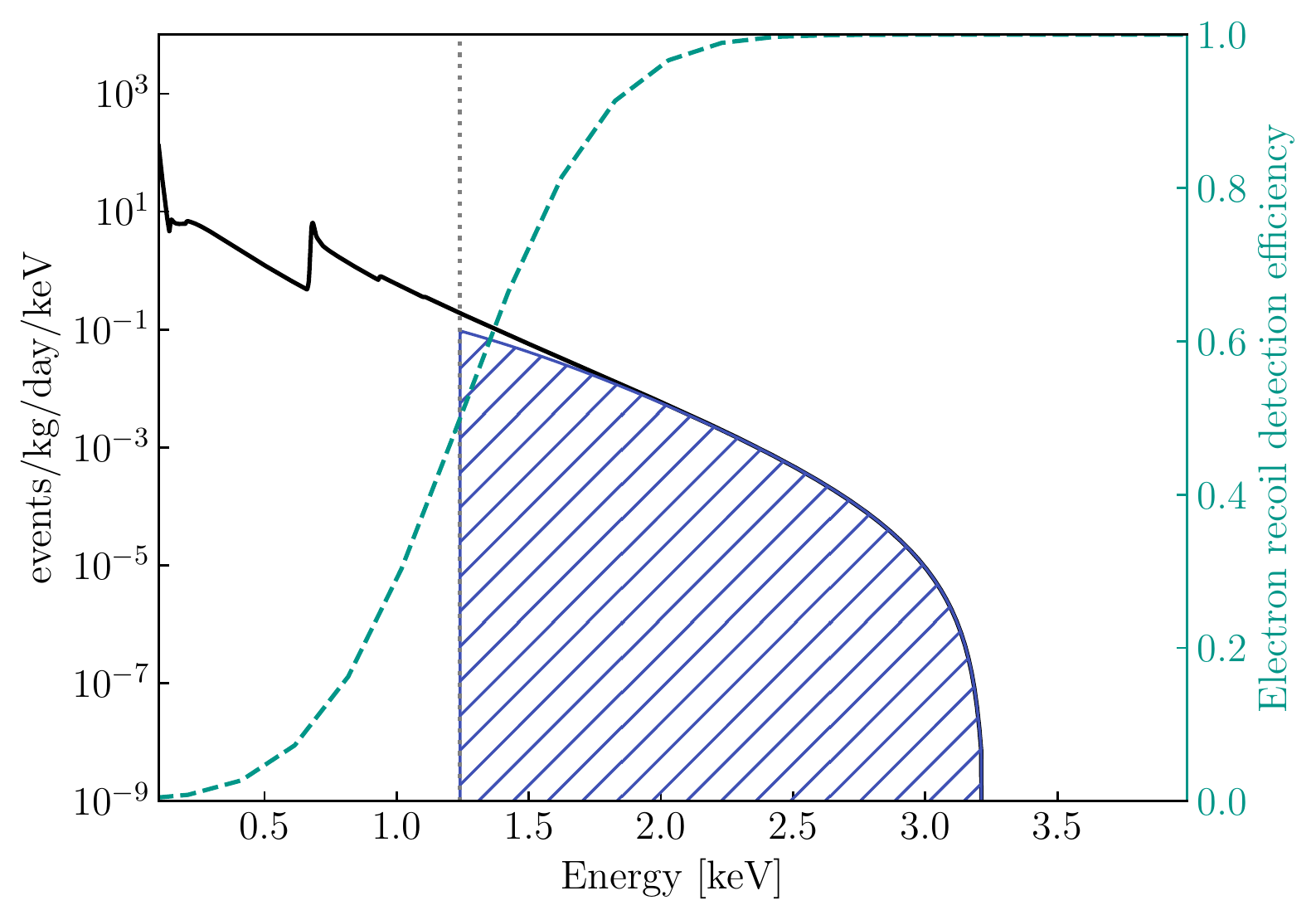}
\caption{Illustration of the DM-nucleus scattering event rate from the Migdal effect with a heavy scalar mediator (solid black line) for $m_{\mathrm{DM}}=1$~GeV/c$^2$ with a cross section per nucleus of $1\times10^{-35}$~cm$^2$. The scattering event rate was calculated following Ref.~\cite{Ibe:2017yqa}. Also shown is the efficiency from the \textit{in situ} tritium measurements performed by the LUX detector (dashed teal line). The hatched blue area indicates the event rate considered for this analysis with tritium efficiency and a 1.24~keV energy threshold (dotted gray line) applied. Data quality cuts are not included.}
\label{fig:signal_rate}
\end{figure}

\begin{figure}[b!]
\includegraphics[width=85mm]{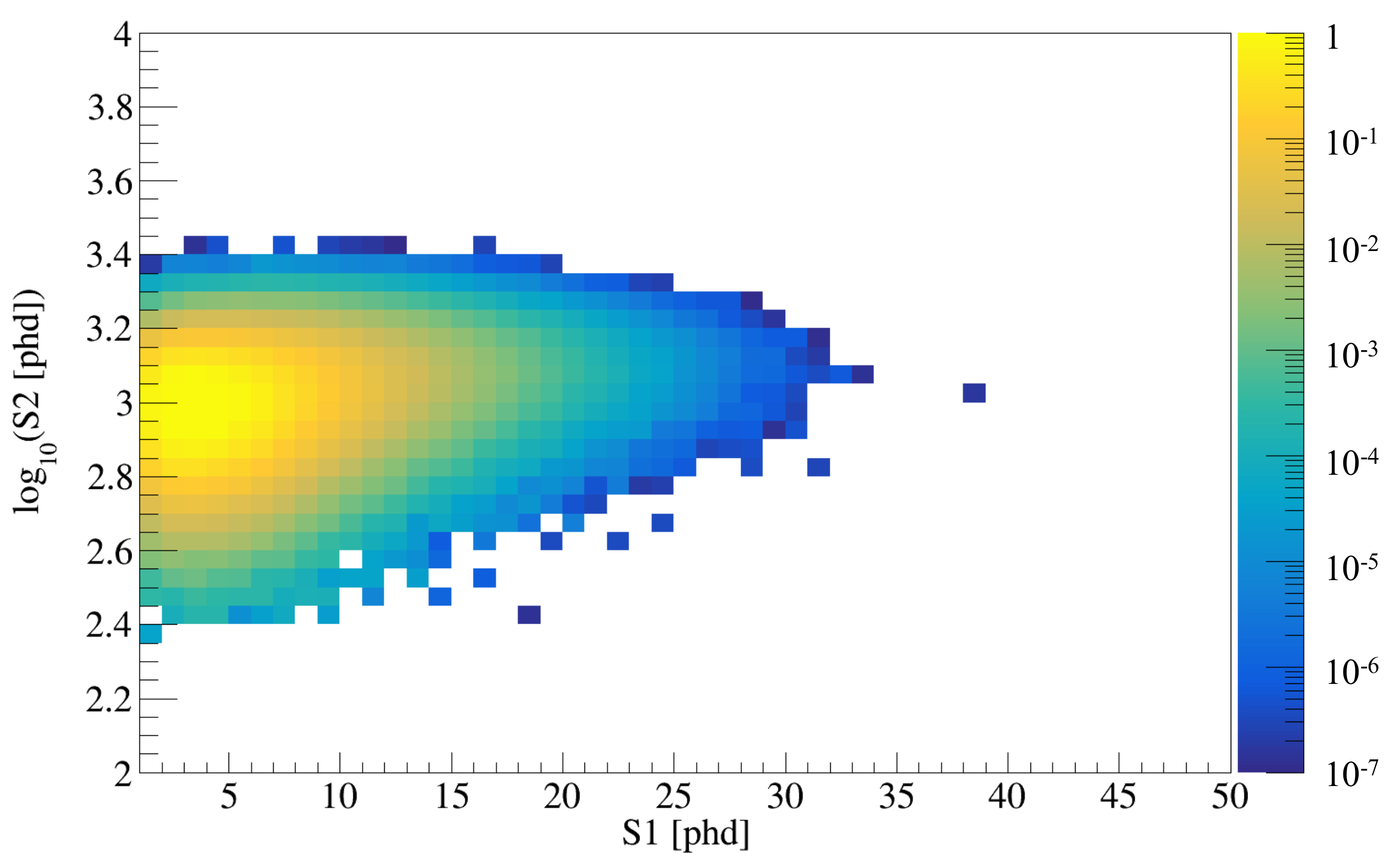}
\caption{The expected signal from DM-nucleus interactions through the Migdal effect with a cross section per nucleus of $1\times10^{-35}$~cm$^2$ projected onto a two-dimensional space of log$_{10}S2$ vs. $S1$. Assumptions are the same as in Fig.~\ref{fig:signal_rate} with additional data quality cuts applied.}
\label{fig:S1S2_signal}
\end{figure}

The expected event rate for a 1~GeV/c$^2$ DM particle with a cross section per nucleus of $1\times10^{-35}$~cm$^2$, the detector ER efficiency, and the low-energy cutoff are illustrated in Fig.~\ref{fig:signal_rate}. The resulting signal model projected on the two-dimensional space of $S1$-log$_{10}S2$ with all analysis cuts applied is shown in Fig.~\ref{fig:S1S2_signal}.

\textit{Background model.---}An important distinction between WS2013 and this Letter is that the sub-GeV signal from both the Bremsstrahlung and Migdal effects would result in additional events within the ER classification, as identified by the ratio of $S2$ to $S1$ size. The standard WIMP search only has a small background from leakage of ER events into the NR band. However, both the sub-GeV signal and most backgrounds are in the ER band, so ER-NR discrimination cannot be used to reduce backgrounds in this analysis. The ER band is populated significantly, with contributions from $\gamma$-rays and $\beta$ particles from radioactive contamination within the xenon, detector instrumentation, and external environmental sources as described in~\cite{Akerib:2014rda}. For further information about the background model, refer to~\cite{Akerib:2015rjg,Akerib:2017vbi} as the background model used in this Letter is identical.

\begin{figure}[t!]
\includegraphics[width=85mm]{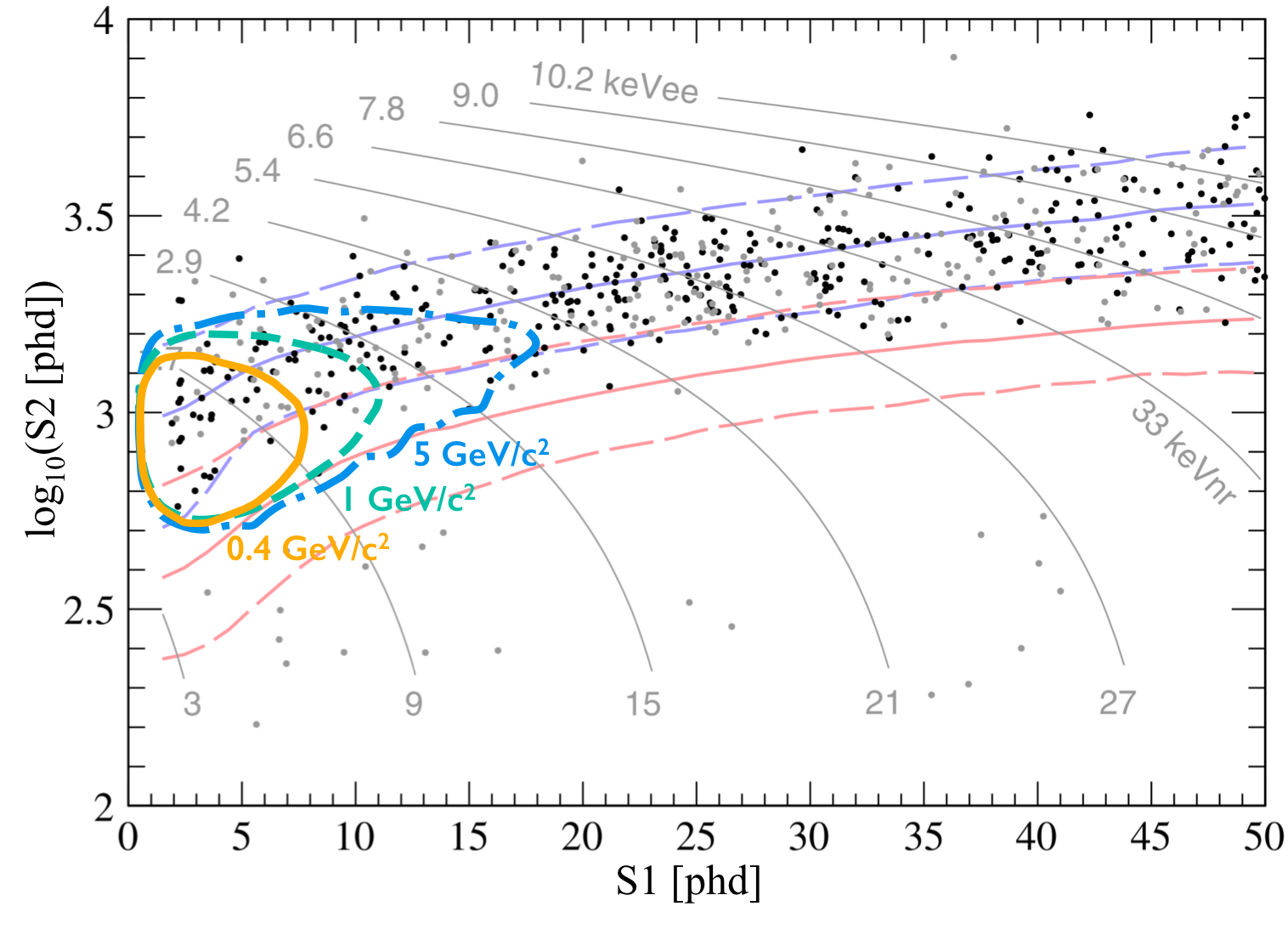}
\caption{Contours containing 95\% of the expected DM signal from the Bremsstrahlung and Migdal effects using NEST package v2.0~\cite{nest2.0}. The solid amber contour indicates a Bremsstrahlung signal of $m_{\mathrm{DM}}=0.4$~GeV/c$^2$ assuming a heavy scalar mediator (7.9 events). The other two contours are for the Migdal effect: The dashed teal contour is for $m_{\mathrm{DM}}=1$~GeV/c$^2$ assuming a heavy scalar mediator (10.8 events), and the dash-dot light blue contour is for $m_{\mathrm{DM}}=5$~GeV/c$^2$ assuming a light vector mediator (11.5 events). The number in parentheses indicates the expected number of signal events within the contour for a given signal model with a cross section at the 90\% C.L. upper limit. The contours are overlaid on 591~events observed in the region of interest from the 2013 LUX exposure of 95~live days and 145~kg fiducial mass (cf. Ref~\cite{Akerib:2015rjg}). Points at radius $<18$~cm are black; those at 18-20~cm are gray since they are more likely to be caused by radio contaminants near the detector walls. Distributions of uniform-in-energy electron recoils (blue) and an example signal from $m_{\mathrm{DM}}=$50~GeV/c$^2$ (red) are indicated by 50th (solid), 10th, and 90th (dashed) percentiles of $S2$ at given $S1$. Gray lines, with an ER scale of keVee at the top and Lindhard-model NR scale of keVnr at the bottom, are contours of the linear-combined $S1$-and-$S2$ energy estimator~\cite{Shutt:2007zz}.}
\label{fig:signal_example}
\end{figure}

\begin{figure}[t!]
\includegraphics[width=85mm]{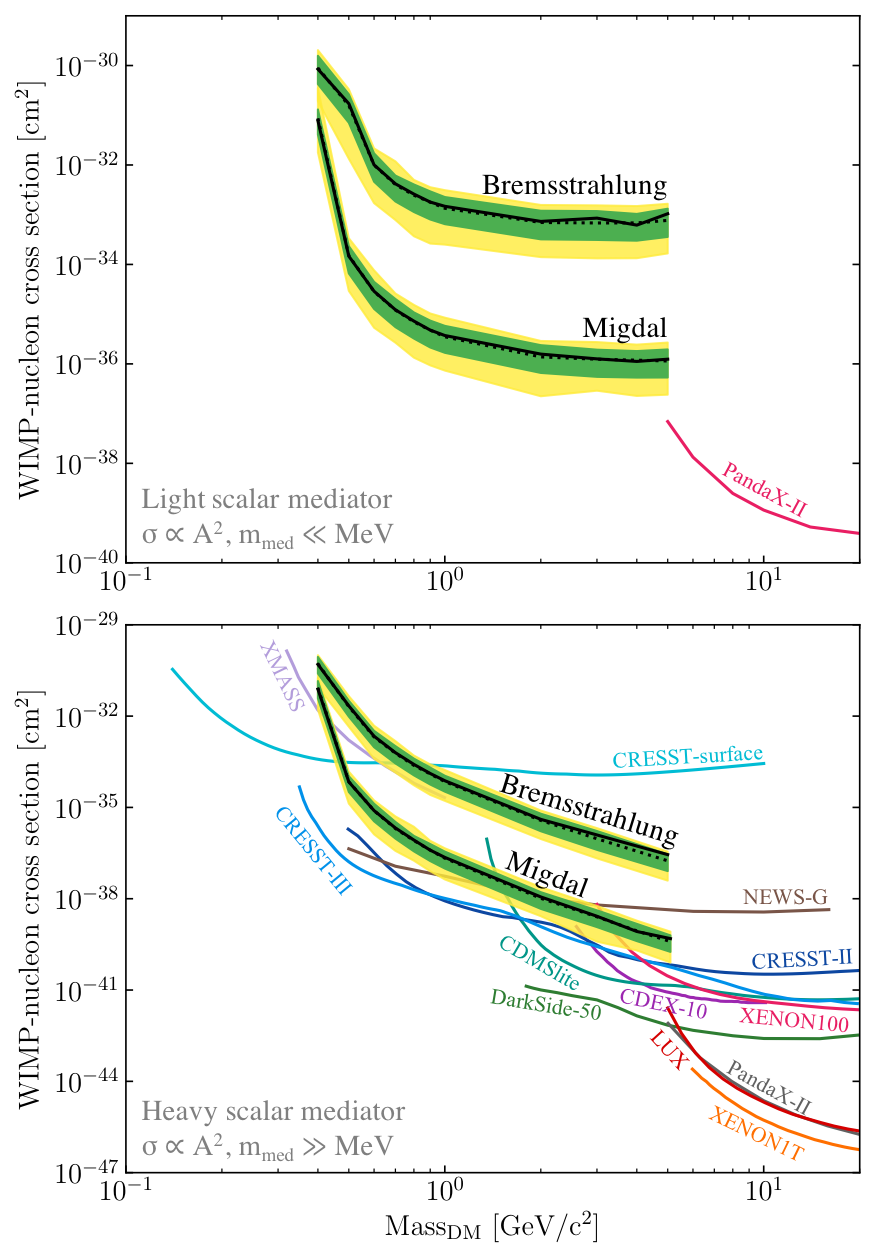}
\caption{Upper limits on the SI DM-nucleon cross section at 90\% C.L. as calculated using the Bremsstrahlung and Migdal effect signal models assuming a scalar mediator (coupling proportional to $A^2$). The 1- and 2-$\sigma$ ranges of background-only trials for this result are presented as green and yellow bands, respectively, with the median limit shown as a black dashed line. The top figure presents the limit for a light mediator with $q_\mathrm{ref}=1$~MeV. Also shown is a limit from PandaX-II~\cite{Ren:2018gyx} (pink), but note that Ref.~\cite{Ren:2018gyx} uses a slightly different definition of $F_\mathrm{med}$ in their signal model. The bottom figure shows limits for a heavy mediator along with limits from the SI analyses of LUX~\cite{Akerib:2016vxi} (red), PandaX-II~\cite{Cui:2017nnn} (gray), XENON1T~\cite{xenon1T} (orange), XENON100 S2-only~\cite{Aprile:2016wwo} (pink), CDEX-10~\cite{Jiang:2018pic} (purple), CDMSlite~\cite{Agnese:2015nto} (teal), CRESST-II~\cite{Angloher:2015ewa} (dark blue), CRESST-III~\cite{Petricca:2017zdp} (light blue), CRESST-surface~\cite{Angloher:2017sxg} (cyan), DarkSide-50~\cite{Agnes:2018ves} (green),   NEWS-G~\cite{Arnaud:2017bjh} (brown), and XMASS~\cite{Kobayashi:2018jky} (lavender).}
\label{fig:limit_scalar}
\end{figure}

\textit{Results.---} The sub-GeV DM signal hypotheses are tested with a two-sided profile likelihood ratio (PLR) statistic. For each DM mass, a scan over the SI DM-nucleon cross section is performed to construct a 90\% confidence interval, with the test statistic distribution evaluated by Monte Carlo sampling using the \textsc{RooStats} package~\cite{RooStats}. Systematic uncertainties in background rates are treated as nuisance parameters with Gaussian constraints in the likelihood. Six nuisance parameters are included for low-$z$-origin $\gamma$-rays, other $\gamma$-rays, $\beta$ particles, $^{127}$Xe, $^{37}$Ar, and wall counts, as described in~\cite{Akerib:2015rjg} (cf. Table I). Systematic uncertainties from light yield have been studied but were not included in the final PLR statistic since their effects were negligible. This is expected as the error on light yield obtained from the tritium measurements ranges from $~10\%$ at low energies to sub 1\% at higher energies. Moreover, slightly changing the light yield is not expected to change the limit significantly since only a small fraction of events near the applied energy threshold are affected.

For an illustration of the expected location of the signal in the $S1$-log$_{10}S2$ detector space, contours for various DM masses with different mediators are overlaid on the observed events from WS2013 shown in Fig.~\ref{fig:signal_example}.

Upper limits on cross section for DM masses from 0.4 to 5~GeV/c$^2$ for both the Bremsstrahlung and Migdal effects assuming both a light and a heavy scalar mediator are shown in Fig.~\ref{fig:limit_scalar}. Upper limits for a light and a heavy vector mediator for the Migdal effect were also calculated. The limits are scaled by $Z^2/A^2$ compared to the scalar mediator case and can be found in~\cite{tvrznikova2019}. The observed events are consistent with the expectation of the background-only hypothesis.

\textit{Summary.---}Contributions from the Bremsstrahlung and Migdal effects extend the reach of the LUX detector to masses previously inaccessible via the standard NR detection method. The Bremsstrahlung photon and the electron from the Migdal effect emitted from the recoiling atom boost the scattering signal for low mass DM particles since the energy transfer is larger in these atomic inelastic scattering channels than in the standard elastic channel and the ER efficiency is significantly higher at low energies. This analysis places limits on SI DM-nucleon cross sections to DM from 0.4~GeV/c$^2$ to 5~GeV/c$^2$ assuming both scalar and vector, and light and heavy mediators. The resulting limits achieved using the Migdal effect, in particular, create results competitive with detectors dedicated to searches of light DM. Furthermore, this type of analysis will be useful to the next-generation DM detectors, such as LZ~\cite{Akerib:2018lyp} by extending their reach to sub-GeV DM masses.

\textit{Acknowledgments.---}The authors would like to thank Rouven Essig, Masahiro Ibe, Christopher McCabe, Josef Pradler, and Kathryn Zurek for helpful conversations and correspondence. We would also like to thank the referees for their constructive comments and recommendations.

This Letter was partially supported by the U.S. Department of Energy (DOE) under Award No. DE-AC02-05CH11231, DE-AC05-06OR23100, DE-AC52-07NA27344, DE-FG01-91ER40618, DE-FG02-08ER41549, DE-FG02-11ER41738, DE-FG02-91ER40674, DE-FG02-91ER40688, DE-FG02-95ER40917, DE-NA0000979, DE-SC0006605, DE-SC0010010, DE-SC0015535, and DE-SC0019066; the U.S. National Science Foundation under Grants No. PHY-0750671, PHY-0801536, PHY-1003660, PHY-1004661, PHY-1102470, PHY-1312561, PHY-1347449, PHY-1505868, and PHY-1636738; the Research Corporation Grant No. RA0350; the Center for Ultra-low Background Experiments in the Dakotas (CUBED); and the South Dakota School of Mines and Technology (SDSMT). Laborat\'{o}rio de Instrumenta\c{c}\~{a}o e F\'{i}sica Experimental de Part\'{i}culas (LIP)-Coimbra acknowledges funding from Funda\c{c}\~{a}o para a Ci\^{e}ncia e a Tecnologia (FCT) through the Project-Grant PTDC/FIS-NUC/1525/2014. Imperial College and Brown University thank the UK Royal Society for travel funds under the International Exchange Scheme (IE120804). The UK groups acknowledge institutional support from Imperial College London, University College London and Edinburgh University, and from the Science \& Technology Facilities Council for Grants ST/K502042/1 (AB), ST/K502406/1 (SS), and ST/M503538/1 (KY). The University of Edinburgh is a charitable body, registered in Scotland, with Registration No. SC005336.\\
This research was conducted using computational resources and services at the Center for Computation and Visualization, Brown University, and also the Yale Science Research Software Core.\\
We gratefully acknowledge the logistical and technical support and the access to laboratory infrastructure provided to us by SURF and its personnel at Lead, South Dakota. SURF was developed by the South Dakota Science and Technology Authority, with an important philanthropic donation from T. Denny Sanford. Its operation is funded through Fermi National Accelerator Laboratory by the Department of Energy, Office of High Energy Physics.

\bibliography{sub_gev}

\end{document}